\documentclass[reprint,amsmath,amssymb,aps]{revtex4-2} 

\usepackage{graphicx}
\usepackage{dcolumn}
\usepackage{bm}
\usepackage{url}
\usepackage{comment}
\usepackage{hyperref}

\begin{document}
\title{Reverse Monte Carlo Modelling in Reduced Dimensions}

\author{Md Golam Moctader}
\affiliation{Department of Physics and Astronomy, University of Turku, FI-20014 Turun yliopisto, Finland}%

\author{Eemeli A. Eronen}
\affiliation{Department of Physics and Astronomy, University of Turku, FI-20014 Turun yliopisto, Finland}%

\author{Johannes Niskanen}
\email{johannes.niskanen@utu.fi}
\affiliation{Department of Physics and Astronomy, University of Turku, FI-20014 Turun yliopisto, Finland}%

\date{\today}
\begin{abstract}
Physical experiments frequently result in inverse problems, where the recorded response is used to deduce information about the sample, typically yielding only partial recovery. This work presents a reverse Monte Carlo modeling method for solving such problems involving statistical averaging, {\it e.g.} at the nanoscale. The proposed algorithm operates in an input subspace of the forward problem, with dimensions reduced from those of the original input space. The dimensionality reduction focuses on features that influence the response, as irrelevant information is lost in the process. We first study the performance of the method in tailored test cases that also reveal information loss when several input characteristics produce similar responses. Finally, we apply the method to a previously published simulated dataset of X-ray absorption, X-ray emission, and X-ray photoelectron spectra of the water molecule.
\end{abstract}

\maketitle
\section{Introduction}

Given a forward problem $\mathbf{y} = f(\mathbf{x})$ for the response variable $\mathbf{y}$ as a function $f$ of the input $\mathbf{x}$, the corresponding inverse problem consists of deducing the input $\mathbf{x}$ from the observed response $\mathbf{y}$.
The task is nontrivial, as the inverse function $f^{-1}$ is typically unknown or may fail to exist. This definition also describes the deduction of information from measurements, and consequently inverse problems arise in nearly every area of science and technology. A concrete example for $\mathbf{y}$ is given by X-ray, UV-visible and IR spectra used for structural ($\mathbf{x}$) characterization of materials.

In the nanoworld, such an observation is typically an average $\mathbf{\bar{y}}$ over several responses due to inevitable statistical distribution of $\mathbf{x}$ that exist during the observation \cite{Allen1987, tuckerman2010}. When atomic coordinates in a structurally diverse system such as liquid are presented as $\mathbf{x}$, many properties of the system follow
\begin{equation}
\label{inverse2}
    \bar{\mathbf{y}} = \int\rho (\mathbf{x}) f(\mathbf{x}) \mathrm{d}V,
\end{equation}
where $\rho$ is the probability distribution of structures, $\mathrm{d}V$ is the volume element of integration, and $f(\mathbf{x})$ is the response given by structure $\mathbf{x}$. While spectroscopy is a widely applied family of techniques, mathematical analysis of the information recovery from observed spectra is quite rare.
\par
The input space may be of notably high dimensionality, which renders the evaluation of the integral (\ref{inverse2}) unfeasible. Instead, structural sampling is typically carried out for an estimate $\left\langle\mathbf{y}\right\rangle$ of the ensemble average $\bar{\mathbf{y}}$. Moreover, the solution to this inverse problem is a distribution, which generally is not a $\delta$-function. To solve this $\rho$, reverse Monte Carlo (RMC) Modelling \cite{mcgreevy1988reverse,McGreevy2001} has been applied. In the method, this distribution is accumulated by sampling averages $\left\langle\mathbf{y}\right\rangle$ of responses $\mathbf{y}$ for a given number of inputs $\mathbf{x}$, which are being varied in steps with probabilities depending on the match with the the observation $\bar{\mathbf{y}}$. Quick evaluation of $f$ is crucial, because the required number of steps is large. Apart from computationally light models and approaches relying on databases of stored values \cite{Wikfeldt2009,Pettersson2022}, a solution may arise from data-driven methods such as machine learning (ML) that rely on quickly approximating the response for the input \cite{penfold2024machine}. 
\begin{figure}[t]
\centering
\includegraphics[width=\columnwidth]{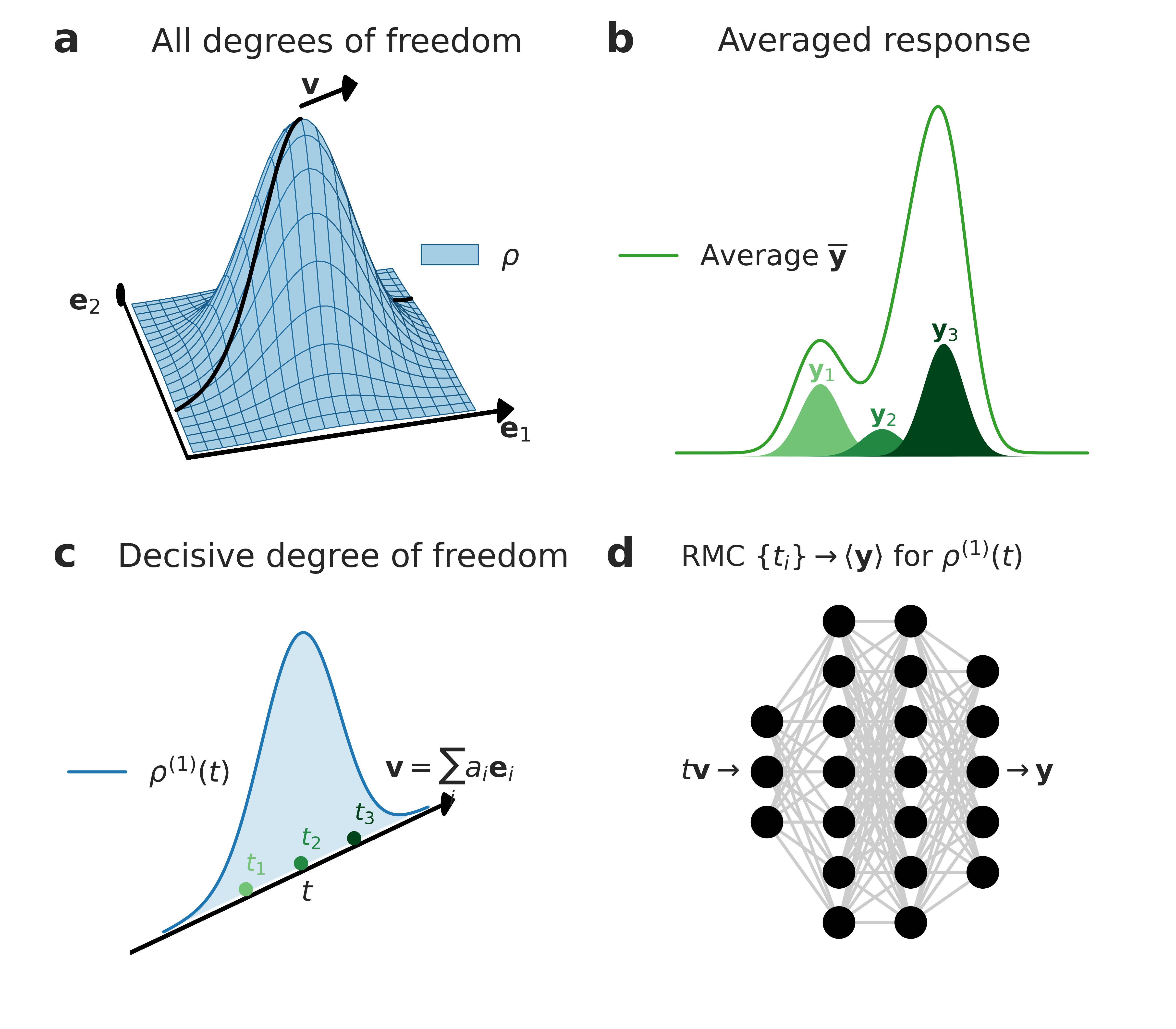}
\caption{\label{fig:principle}The input values $\mathbf{x}$ are (a) distributed over the space by $\rho$ yielding (b) response $\bar{\mathbf{y}}$ as a mean of respective responses $\mathbf{y}$. (c) In this case, the decisive basis vector $\mathbf{v}$ spans a subspace and the corresponding coordinate $t$ dictates the response $\mathbf{y}$ for the data point $\mathbf{x}$. The distribution $\rho^{(1)}$ in the subspace is subject to (d) RMC modeling with the help of numerically fast emulator.}
\end{figure}
\par
The RMC protocol circumvents the need for finding the inverse function, but may still suffer from the high dimensionality of the input. Namely, the procedure will sample every degree of freedom, regardless whether they matter for the response or not. Consequently, a distribution will be obtained also for degrees of freedom that have no effect on the response. Such an analysis is prone to error, as any other distribution along an irrelevant degree of freedom would yield the same averaged response and therefore be an equally likely outcome. For example, X-ray and UV-visible spectra are sensitive to only certain latent variables of the atomistic system \cite{eronen2024structural, eronen2025structural, eronen2025statistics}. Finding these decisive latent degrees of freedom is non-trivial, but promises a tremendous reduction of the dimensionality. 
\par
In this work, we carry out RMC in a subspace of the input domain selected for its significance for explaining the variation of the response. Consequently, the method illustrated in Figure \ref{fig:principle} focuses on information that even {\it a priori} can affect the response. We first test the method with a simple tailored problem with one and two decisive latent variables, and then proceed with an analysis of previously published, simulated data sets of X-ray absorption spectra (XAS), X-ray emission spectra (XES), and X-ray photoelectron spectra (XPS) of the H$_2$O molecule. Satisfactory solution to the presented inverse problem is found when three conditions are met: (i) reconstruction is restricted to the subspace relevant for the variation of the response, (ii) the responses along different latent variables are distinct from each other, and (iii) responses behave monotonously along the latent degrees of freedom.

\section{Methods}
Both phases of the procedure require an ML emulator for the forward problem. We use a neural network (NN), a contemporary machine learning approach, to predict target values from a given input. First, dimensionality reduction is done so that the obtained input subspace covers maximal variation of the response in a given data set, when inputs are projected onto the subspace. We perform this by applying emulator-based component analysis \cite{Niskanen2022a} (ECA). The algorithm utilises an emulator $\mathbf{y}_\mathrm{emu}$ capable of quickly making predictions for the forward problem on new inputs. As a result a basis set $\{\mathbf{v}_i\}_{i=1}^k$ is obtained, yielding an approximate rank-$k$ presentation $\mathbf{x}^{(k)}$ for input $\mathbf{x}$ with a coordinate vector $\mathbf{t}=(t_1,t_2,...,t_k)$
\begin{equation}
\mathbf{x}^{(k)}=\sum_{i=1}^k\underbrace{\left\langle\mathbf{v}_i|\mathbf{x}\right\rangle}_{=:t_i}\mathbf{v}_i.
\label{eq:projection}
\end{equation}
We obtain the vectors $\mathbf{v}_i$ by maximizing variance-weighted $R^2$ score between known target values and ones predicted for the projected data by the emulator. The $R^2$ metric represents the proportion of response variance explained by the model. Letting $\mathbf{Y}_{nm}$ denote the response vector component $m$ of data point $n$, the score is defined as
\begin{equation}
R^2 = 1 - \frac{\sum_{n=1}^{N} \sum_{m=1}^{M}\left(\mathbf{Y}^{(\mathrm{pred})}_{nm} - \mathbf{Y}_{nm}\right)^2}
{\sum_{n=1}^{N} \sum_{m=1}^{M}\left(\mathbf{Y}_{nm} - \left\langle\mathbf{y}\right\rangle_{m}\right)^2}
\label{eq:r2}
\end{equation}
which is maximized with respect to the set of orthonormal basis vectors $\{\mathbf{v}_i\}_{i=1}^{k}$. In the above, $N$ denotes the total number of samples and $M$ the total number of response variables. The elements $\mathbf{Y}_{nm}^{(\mathrm{pred})}$ denote the predicted responses corresponding to the projected inputs $\{\mathbf{x}^{(k)}\}$, while $\mathbf{Y}_{nm}$ denote the corresponding known responses. The term $\left\langle\mathbf{y}\right\rangle_{m}$ represents the mean of the known responses in column $m$, {\it i.e.} averaged over all samples $n$. The numerator thus corresponds to the total squared prediction error, whereas the denominator represents the total variance of the data. The ECA method assumes that the input vectors $\mathbf{x}$ are z-score standardized prior to analysis.
\par
The proposed dimensionality-reduced reverse Monte Carlo (DRRMC) modelling focuses on sampling $N_s$ vectors $\mathbf{t}^{(n)}$ ($n=1,...,N_s$) for an ECA decomposition (\ref{eq:projection}) to establish their distribution for a given reference response $\left\langle\mathbf{y}\right\rangle$. One of these vectors is varied at a time for a trial step with 
\begin{equation}
\chi^2 = \frac{\left(\left\langle\mathbf{y}\right\rangle -\frac{1}{N_s}\sum_{n=1}^{N_s}\mathbf{y}_\mathrm{emu}\left(\sum_it_i^{(n)}\mathbf{v}_i\right) \right)^2}{\sigma^2}
\end{equation}
being evaluated using a pre-determined parameter $\sigma$. The trial step is accepted with the probability $\min\{1,\mathrm{e}^{-\Delta\chi^2}\}$, where $\Delta\chi^2$ is the change in the $\chi^2$ value during the step. The distribution of the coordinate vectors is accumulated as a histogram with a given step interval. As the final result of the procedure, a reconstructed distribution $\rho^{(k)} (\mathbf{t})$ in the $k$-dimensional ECA space is obtained. Using Equation (\ref{eq:projection}) this result can be converted back to a subspace of the inputs, with the note that the sign of a coordinate $t_i$ and that of the respective vector $\mathbf{v}_i$ of Equation (\ref{eq:projection}) can be changed together to represent the same point $\mathbf{x}^{(k)}$. We constrain the method to only sample the components of vectors $\mathbf{t}$ within the region covered by the training data. This choice inherently limits the analysis to the realm of expected validity of NN emulation, but does not bias it with additional information a more restrictive choice would give to the algorithm. For this reason, the training dataset must include all the relevant conditions the model is expected to handle during prediction. 
\par
We combine DRRMC with a \texttt{Pytorch} \cite{Paszke2019} powered implementation \cite{vladyka2024implementation} of the ECA algorithm. Along with the number of sampling points $N_s$, the DRRMC implementation offers an option to adjust $\sigma$ until a desired nominal step acceptance probability is achieved, after which sampling is logged. We use the ML training data to select the regions of notable (1\% of maximum) response for ECA decomposition. We then test reconstruction of $\rho^{(k)}$ using a separate test set unseen by the earlier stages of the procedure. The known inputs in the test set are used only for assessment of the results of the analysis.

\section{Results}
We first study this method by using tailored data sets with one and two decisive latent variables. For training data of NN, we sampled 100\,000 data points $(\mathbf{x}, \mathbf{y})$ so that each component for the $\mathbf{x}$ data were independent following a uniform distribution over the range [-1,1]. All the trained emulators had $R^2$ score better than 0.99 in cross-validation. Previous works have shown the dimensionality of a reasonable input to be in the order of hundreds to a few thousands \cite{eronen2024structural}, and therefore we chose the dimensionality of the input to be 512. To mimic a workflow of a real case, we further z-score standardized all input data using a scaler established by the according training set. To investigate information recovery by the method, we first study a case where the response is dictated by one latent variable, and then proceed to a case of two relevant latent variables. Last, we will investigate a simulated data set for the H$_2$O molecule from Ref. \citenum{Niskanen2022a}.
\par
Information loss can be seen in a multidimensional case, where the target outcome is dictated by only one latent degree of freedom $t=\mathbf{v}\cdot\mathbf{x}$ with a predefined $\mathbf{v}$. As the forward function we use a vector-valued $\mathbf{y}(\mathbf{x})$ returning a gaussian centered at position $0.5+0.1t$ on a grid $\mathbf{g}$ ranging from 0 to 1.0. Such a response could represent {\it e.g.} a core-level photoelectron spectrum of an element over a range of core level binding energy values, whereas $\mathbf{x}$ would then be associated with the molecular structure. As a single degree of freedom in input is decisive for the position of the peak on the grid, the first ECA component $\mathbf{v}_1=\mathbf{v}$ will cover 100\% of target variation. Consequently, only $t_1$ has an effect on the response, and therefore its distribution $\rho^{(1)}$ is the only reconstructable information from the inverse problem. The presented example is not unrealistic: for example simulated O K-edge binding energy of an isolated H$_2$O molecule depends nearly completely on the H-O-H angle \cite{Niskanen2022a}. For the dimensionality reduction step, we performed the ECA fit using the training data after z-score standardization.
\par
\begin{figure}
    \centering
    \includegraphics[width=\linewidth]{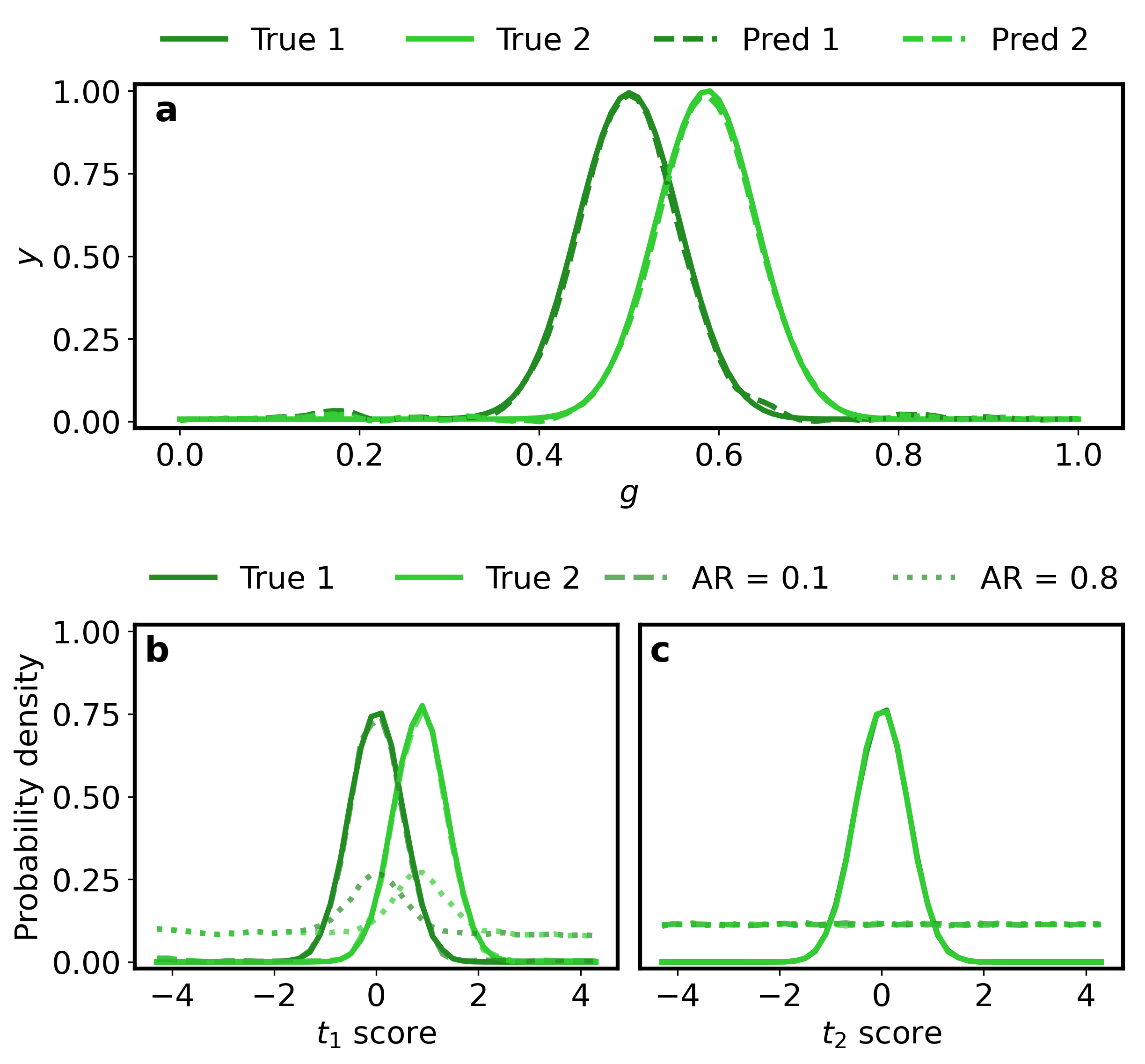}
    \caption{Reconstruction with 512-dimensional input data with one decisive latent variable $t_1$. (a) The known target $\left\langle\mathbf{y}\right\rangle$ for a set of 100\,000 points shown together with the last sampled target means with acceptance probability 0.1. (b)  Distribution of the $t_1$ scores and their cumulatively sampled reconstruction with acceptance probabilities 0.1 and 0.8 together with the distribution of the known scores for the 100\,000 data points used for $\left\langle\mathbf{y}\right\rangle$.(c) Distribution of the $t_2$ scores and their cumulatively sampled reconstruction with acceptance probabilities 0.1 and 0.8. Due to loss of information, the fit is unable to capture the distribution of $t_2$, as the score has no effect on the response. Note that a 2D fit produces the proper distribution for $t_1$ because $t_2$ is completely decoupled from the problem. For details, see text.} 
    \label{fig:CaseAresults}
\end{figure}

We sample two test sets from (512-dimensional) normal distribution $\rho$, shifted slightly along $\mathbf{v}$. Figure \ref{fig:CaseAresults}a shows the respective reference response $\left\langle\mathbf{y}\right\rangle$ as a mean of the 100\,000 samples in the nearly normally distributed test sets, manually restricted in their input-space coverage to that of the training set. In addition, the figure depicts the response at the last iteration, which shows deviations from $\left\langle\mathbf{y}\right\rangle$, characteristic to the perpetual oscillations within the method. Figure \ref{fig:CaseAresults}b depicts the underlying distribution of the 100\,000 $t_1$ scores, and its reconstruction from $\left\langle\mathbf{y}\right\rangle$ by sampling of 10 million steps with sampling interval 10\,000, with nominal acceptance probabilities 0.1 and 0.8, and using a set of $N_s$=128 one-dimensional vectors $\mathbf{t}$. Information loss is seen when trying reconstruction of $t_2$, here corresponding to any vector $\mathbf{v}_2$ orthogonal to $\mathbf{v}_1$. Figure \ref{fig:CaseAresults}c shows a futile attempt of reconstructing $t_2$, irrelevant for the response in this case.
\par
\begin{figure}
    \centering
    \includegraphics[width=\linewidth]{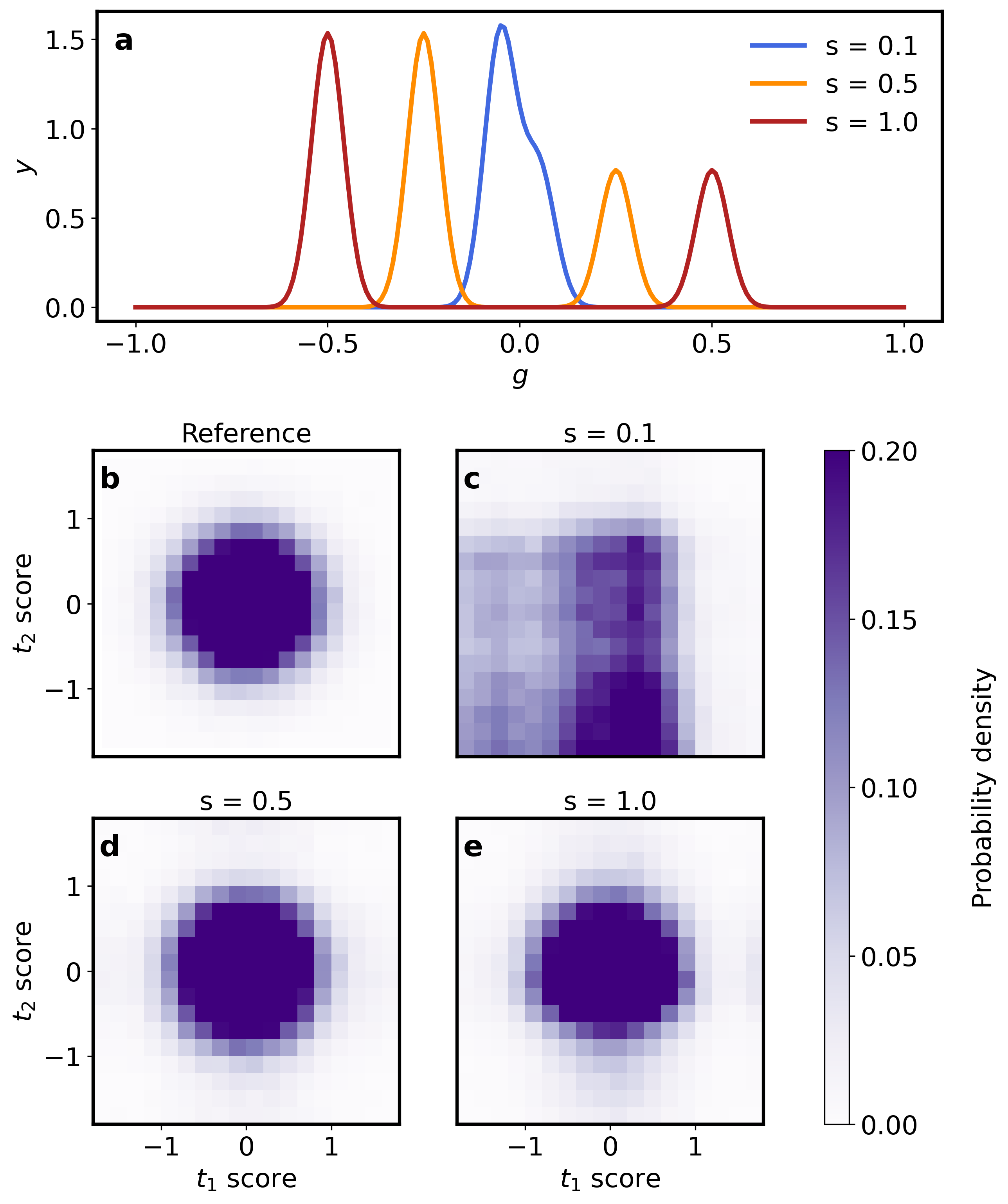}
    \caption{Reconstruction of two decisive latent variables $t_1$ and $t_2$ with acceptance probability 0.1. (a) The reference response $\left\langle\mathbf{y}\right\rangle$ for each mean peak distance $s$. (b) The reference distribution of $t_1$ and $t_2$ and (c-e) the reconstructed distributions for increasing $s$. For details, see text.}
    \label{fig:CaseA2results}
\end{figure}

Next, we turn to simultaneous reconstruction of two scores $\mathbf{t}=(t_1,t_2)$ using the same principles as in the one dimensional case. These cases can be visualized by response function $\mathbf{y}$ of two gaussians evaluated on a grid from -1 to 1. The positions of the two peaks were $-0.5s+0.07t_1$ and $0.5s+0.07t_2$ for $t_1=\mathbf{v}\cdot\mathbf{x}$ and $t_2=\mathbf{v}'\cdot\mathbf{x}$, with $\mathbf{v}\cdot\mathbf{v}'=0$. The two peaks are separated by a distance $s$ on the grid, and we systematically investigate reconstruction quality as this displacement decreases, leading to potential peak overlap. Because sampling is decided blindly in the minimum--maximum range of the train data $t$ scores (rounded wider to one decimal), the parameter value $s=0.1$ presents the case in which the two peaks can be swapped in position, and therefore compensate for each other within the sampling interval. Figure \ref{fig:CaseA2results} depicts the reconstructed distributions in three cases. For the case with smallest $s$, identifying the variable responsible for an effect becomes impossible. This problem is mitigated when the responses for the two latent variables are distinct for all relevant $\mathbf{x}$ as for $s=0.5,1.0$, with two completely separated peaks.
\par
\begin{figure}
    \centering
    \includegraphics[width=\linewidth]{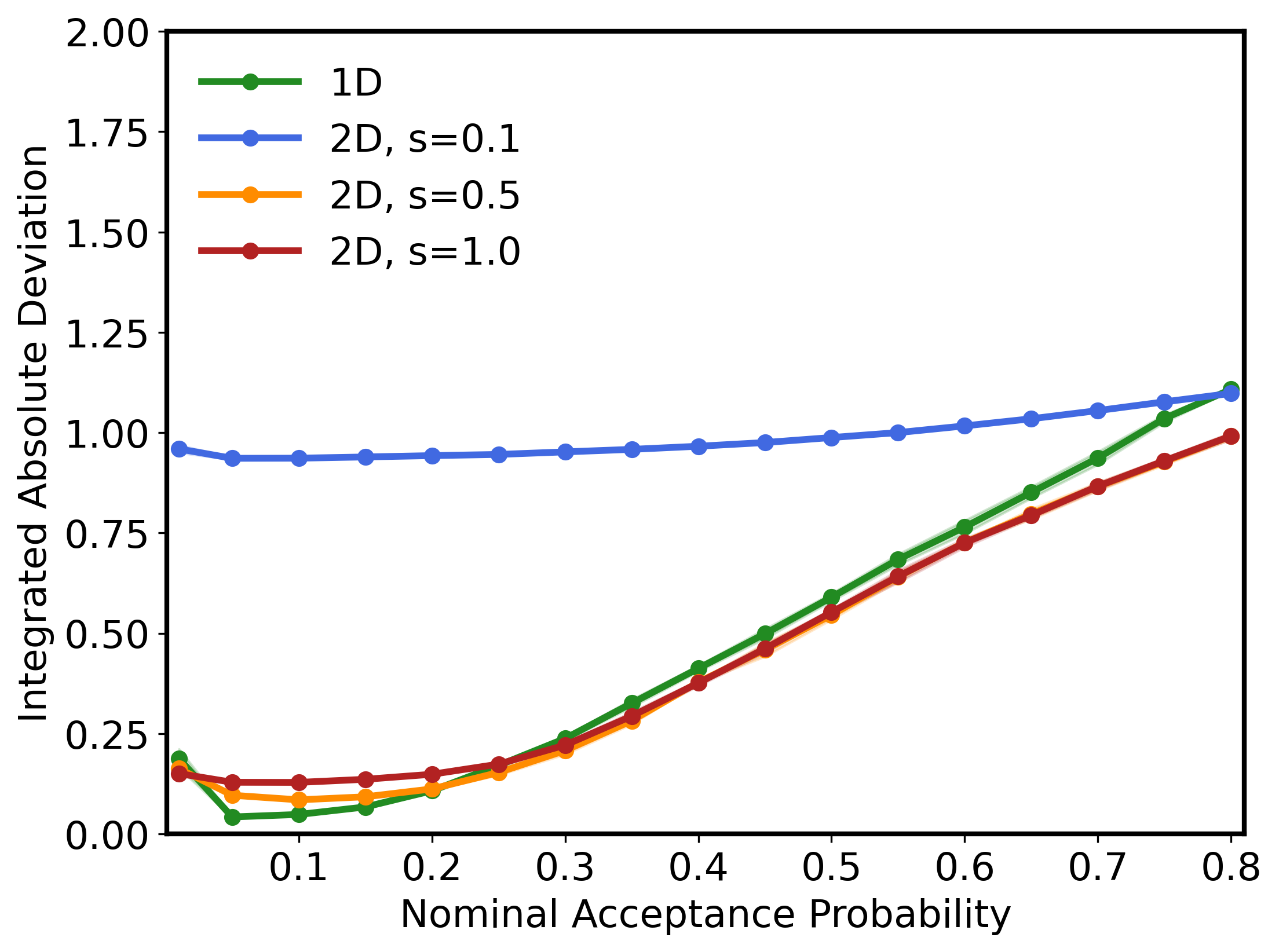}
    \caption{Mean integrated absolute deviation in between the reconstructed and the reference distributions as a function of the nominal acceptance probability in the DRRMC for one- and two-dimensional cases. Curves represent averages over 10 sampling runs with different seeds for the involved random number generators. The shaded regions denote $\pm$1 standard deviation. Since the reference $\left\langle\mathbf{y}\right\rangle$ was evaluated from a large but finite set of 100\,000 points, minor discrepancies can be attributed to statistical noise.    \label{fig:Acceptrate}}
\end{figure}

The method has an inherent parameter $\sigma$, for which a value needs to be selected. We investigated the accuracy of reconstruction as a function of the step-acceptance probability as integrated absolute deviation to the known distribution, depicted in Figure \ref{fig:Acceptrate}. The curves indicate that very low values hinder convergence and limit exploration of the parameter space. High values, in turn, allow for trial steps with averaged response divergent to the reference. This reduces selectivity and degrades accuracy. For the four cases studied thus far, the optimal choice is around 0.1. Since the reference $\left\langle\mathbf{y}\right\rangle$ was evaluated from a large but finite set of inputs $\mathbf{x}$, minor discrepancies may be caused by statistical noise. We studied ten runs with different seeds of the involved random number generators and observed no notable deviation in the accuracy measured using the metrics. Likewise, doubling or halving $N_s$ from the used 128, does not affect the results notably, as demonstrated in the SI.
\par
\par
\begin{figure*}
    \centering
    \includegraphics[width=0.8\linewidth]{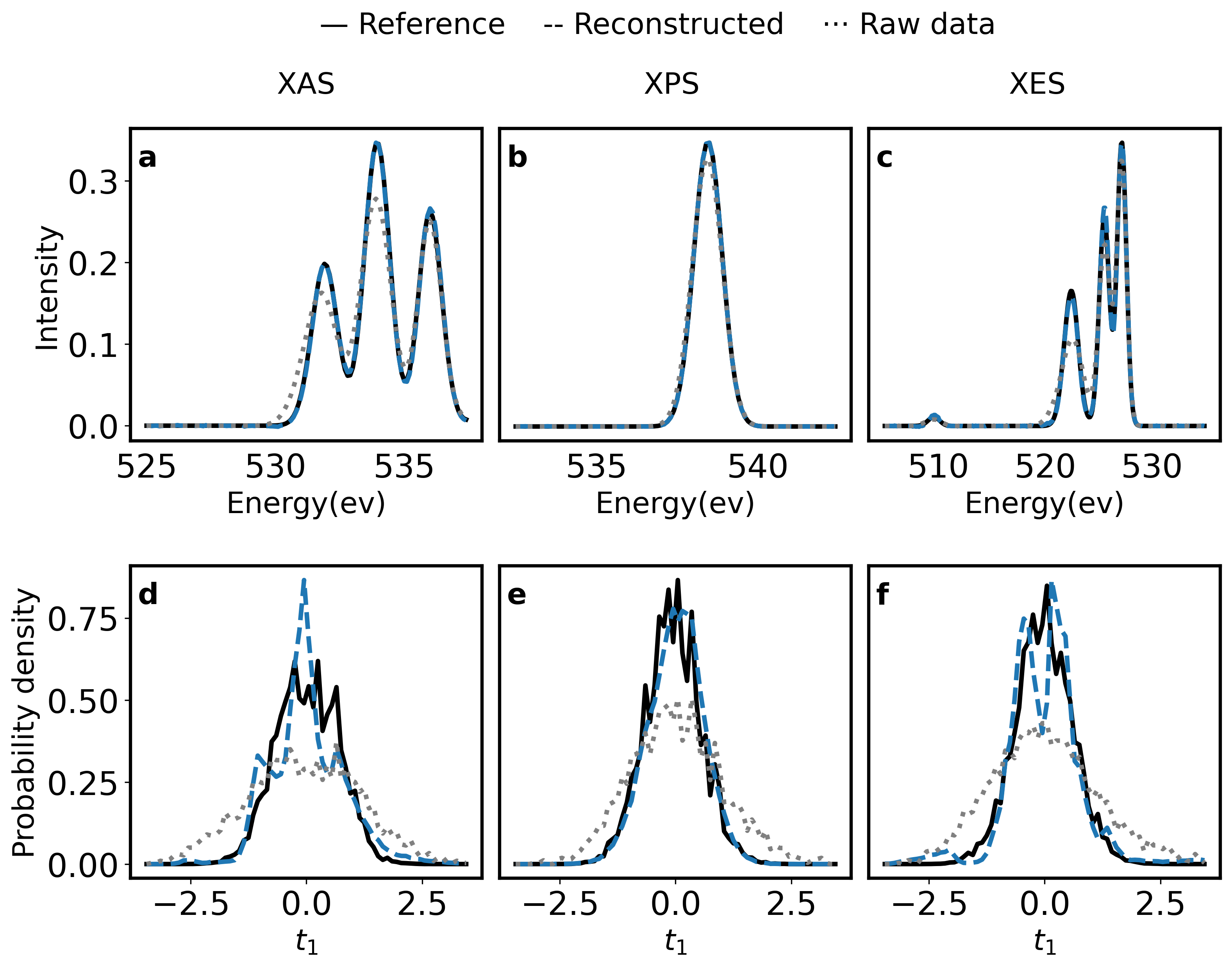}
    \caption{(a–c): Comparison of reference and the last iteration for XAS, XPS, and XES spectra of the H$_2$O molecule in the DRRMC sampling, respectively. (d–f) Comparison of according structural distributions. The raw data is given for comparison. \label{fig:CaseB}}
\end{figure*}
\par
Next, we consider the reconstruction of latent variables for simulated sets of XAS, XES, and XPS for the H$_2$O molecule. These spectra reflect transitions in between O 1s core orbital, valence orbitals and the electronic continuum. Consequently, the spectra are sensitive to the local atomistic structure, which makes these spectroscopic methods useful for structural characterization of materials. The dataset used in this study is adopted from Ref.~\citenum{Niskanen2022a}. We narrow the structural distribution further by weighting the contribution of data points by a narrower gaussian. Thus, the reference to be constructed falls within the training data set and the underlying structural distribution differs from that of the raw data set.
\par
The reference spectra depicted in Figures \ref{fig:CaseB} a--c represent features characteristic of each process, discretized over the relevant energy ranges to yield modality‑specific response vectors to perform DRRMC fitting for. The respective distributions $\rho^{(1)}$ of the latent variable $t_1$ are depicted in Figures \ref{fig:CaseB} d-–f. In these figures, the spectra and distributions of the raw data set are included for visual comparison. The structural information in the dataset is represented using three physically meaningful geometric descriptors: bond angle (degrees), long  bond  (\AA), and short bond (\AA). These variables define the underlying molecular configuration and serve as the input feature space for the reconstruction, to be further reduced in dimensionality by ECA. To quantify the extent to which the spectral variability can be captured within a latent representation, we evaluate the variance explained by the one-dimensional ECA projection for each spectroscopy. The results indicate that this representation captures $68.0\%$ of the generalized variance ($R^2$ score) for XAS, $74.2\%$ for XES, and $98.5\%$ for XPS.
\par
Although only one latent degree of freedom is considered, the sample size $N_s = 128$ allows for faithful representation of the reference spectrum as indicated by the spectra of the last evaluation depicted Figures \ref{fig:CaseB} a--c. This observation manifests loss of structural information in the spectroscopic process, when an average over a distribution is considered.
\par
The reconstructed one-dimensional distrbutions $\rho^{(1)}$ match well with the corresponding reference, presented for $t_1$ along vector $\mathbf{v}_1$ respective to each case. Comparison with the raw data reveals that the changes induced to the spectrum by alteration of this distribution are truthfully captured by one-dimensional DRRMC. However, the three spectra have different relation to the underlying structure, which has consequences for reconstruction of multidimensional distributions. In particular, the associated basis vectors $\mathbf{v}_i$ differ between the three cases, reflecting distinct underlying latent structures in each spectrum. The supporting information depicts results for runs for $\rho^{(2)}$, which notably fail for XPS. This is understood by the spectrum being virtually explained by one latent degree of freedom. Notable deviations from the reference take place for XAS and XES as well.

\section{Discussion}
Whereas information irrelevant for the response is irrecoverable, the tailored examples show that the decisive information may be reconstructed. A similar observation has been made by Vladyka {\it et al.}\cite{vladyka2023towards} and Timoshenko {\it et al.} \cite{timoshenko2017supervised} for reconstruction of certain characteristics from spectroscopic probes. For an extensive review we refer to Penfold {\it et al.} \cite{penfold2024machine}. In the proposed framework, the problem of insensitivity of the spectral response to certain input features is systematically solved by ECA. By considering a subspace of the input space, the method seeks for an injective approximate forward function. Even though the full input point can be represented with expansion (\ref{eq:projection}) done to its full rank, reconstruction of structural information by higher-rank components becomes increasingly unreliable due to the dimishing relevance of these coordinates for the response. However, when different latent variables share an indistinguishable response, knowing the correct cause for an observation becomes impossible even with dimensionality reduction. Consider, for example, a case where two latent variables cause similar shifting of a spectral line. Furthermore, non-monotonous behavior, such as sinusoidal oscillation, along any latent degree of freedom renders identifying the correct input value within the range impossible.
\par
While the approach may be effective in theoretical simulations, its blind application to experimental data may lead to erroneous results, as correctness of simulations and accurate signal fidelity are critical for reliable interpretation. Consequently, careful response feature selection together with accurate emulation may prevent amplification of irrelevant or noisy components. Thus a potentially safer way for interpration is given by analysis of simulations that reproduce an experimentally observed effect. Another promising strategy to mitigate the risk is the use of regions of interest (ROIs), which concentrate analysis on physically meaningful response segments while excluding areas susceptible to noise, or prone to overinterpretation of small details due to inaccuries of the model. However, coarsening to ROI intensities is a form of information loss in itself. 
\par
One way to resolve these problems is to carry experiments with a separate methods for a set of responses that might yield complementary information, and merge the results into a combined response vector. Therefore, an argument can be made to use DRRMC for investigation of information recovery from the merged response, with which the rank of reliable reconstruction would be higher. Furthermore, applying this method to more complex systems than studied here would require larger datasets, more sophisticated structural descriptors, and possibly new emulator architectures to adequately capture the increased complexity. Still, information recovery is most feasible when the analysis is limited to the degrees of freedom that are most significant for the response.

\section{Conclusions}
We have presented a systematic formalism utilizing machine learning for information recovery from lossfull inverse problems of the integral type (\ref{inverse2}), occuring for example in interpretation of probe signals subject to ensemble averaging. The proposed formalism accounts for and builds on the realization that not every input feature is recoverable in such a problem. First, we use dimensionality reduction based on relevance for the response to identify the input features that do have an effect. Then, reconstruction of distributions for the according latent variables becomes possible by reverse Monte Carlo modeling with additional conditions. Namely, information loss may also happen due to similar responses along two or more {\it a priori} relevant latent degrees of freedom, or due to non-monotonous behavior along any single one. These formal results provide a structured framework for interpretation of signals that can be shown to vary dominantly along certain input characteristics and to be subject to inherent statistical variation, {\it e.g.} spectra from atomistic systems.

\section{Acknowledgements}
J.N. thanks professors E. Kukk and L.G.M. Pettersson for discussions and acknowledges funding by Research council of Finland via the academy project grant 367978. The authors acknowledge CSC – IT Center for Science, Finland, for computational resources.

\section{Data availability}
The codes new to this work are available in Zenodo together with the data for the rudimentary case. The corresponding data and scripts are available at (https://seafile.utu.fi/d/f419f54c88c44efe963b/) here and will be published in Zenodo if the manuscript is accepted for publication. The data for the H$_2$O molecule are available at Dryad Digital Repository (https://doi.org/10.5061/dryad.dncjsxm1m). For the scripts to run directly, the unzipped H$_2$O dataset is to be placed in the directory \path{CaseB_H2O/H2O/} of this work.

\bibliography{bibliography}

@Book{Tuckerman2010,
  author    = {Mark E. Tuckerman},
  publisher = {Oxford University Press, Oxford},
  title     = {Statistical Mechanics: Theory and Molecular Simulation},
  year      = {2010},
}

@Book{Allen1987,
  author    = {M. P. Allen and D. J. Tildesley},
  title     = {Computer Simulation of Liquids},
  year      = {1987},
  publisher = {Oxford University Press, Oxford}
}

@article{mcgreevy1988reverse,
  title={Reverse Monte Carlo simulation: a new technique for the determination of disordered structures},
  author={McGreevy, RL and Pusztai, L},
  journal={Molecular simulation},
  volume={1},
  number={6},
  pages={359--367},
  year={1988},
  publisher={Taylor \& Francis}
}

@Article{McGreevy2001,
  author   = {R L McGreevy},
  journal  = {Journal of Physics: Condensed Matter},
  title    = {Reverse Monte Carlo modelling},
  year     = {2001},
  month    = {nov},
  number   = {46},
  pages    = {R877},
  volume   = {13},
  doi      = {10.1088/0953-8984/13/46/201}
}

@incollection{Paszke2019,
title = {PyTorch: An Imperative Style, High-Performance Deep Learning Library},
author = {Paszke, Adam and Gross, Sam and Massa, Francisco and Lerer, Adam and Bradbury, James and Chanan, Gregory and Killeen, Trevor and Lin, Zeming and Gimelshein, Natalia and Antiga, Luca and Desmaison, Alban and Kopf, Andreas and Yang, Edward and DeVito, Zachary and Raison, Martin and Tejani, Alykhan and Chilamkurthy, Sasank and Steiner, Benoit and Fang, Lu and Bai, Junjie and Chintala, Soumith},
booktitle = {Advances in Neural Information Processing Systems 32},
editor = {H. Wallach and H. Larochelle and A. Beygelzimer and F. d\textquotesingle Alch\'{e}-Buc and E. Fox and R. Garnett},
pages = {8024--8035},
year = {2019},
publisher = {Curran Associates, Inc.},

}

@Article{Niskanen2022a,
  author   = {Niskanen, J. and Vladyka, A. and Niemi, J. and Sahle, C.J.},
  journal  = {Royal Society Open Science},
  title    = {Emulator-based decomposition for structural sensitivity of core-level spectra},
  year     = {2022},
  number   = {6},
  pages    = {220093},
  volume   = {9},
  doi      = {10.1098/rsos.220093}
 
}

@article{vladyka2024implementation,
  title={Implementation of the emulator-based component analysis},
  author={Vladyka, Anton and Eronen, Eemeli A and Niskanen, Johannes},
  journal={Journal of Computational Science},
  volume={83},
  pages={102437},
  year={2024},
  publisher={Elsevier}
}

@article{vladyka2023towards,
  title={Towards structural reconstruction from {X-ray} spectra},
  author={Vladyka, Anton and Sahle, Christoph J and Niskanen, Johannes},
  journal={Physical Chemistry Chemical Physics},
  volume={25},
  number={9},
  pages={6707--6713},
  year={2023},
  publisher={Royal Society of Chemistry}
}

@article{penfold2024machine,
  title={Machine-learning strategies for the accurate and efficient analysis of x-ray spectroscopy},
  author={Penfold, Thomas and Watson, Luke and Middleton, Clelia and David, Tudur and Verma, Sneha and Pope, Thomas and Kaczmarek, Julia and Rankine, Conor},
  journal={Machine Learning: Science and Technology},
  volume={5},
  number={2},
  pages={021001},
  year={2024},
  publisher={IOP Publishing}
}

@article{eronen2024structural,
  title={Structural descriptors and information extraction from X-ray emission spectra: aqueous sulfuric acid},
  author={Eronen, E.A. and Vladyka, A. and Sahle, Ch. J. and Niskanen, J.},
  journal={Physical Chemistry Chemical Physics},
  volume={26},
  number={34},
  pages={22752--22761},
  year={2024},
  publisher={Royal Society of Chemistry}
}

@article{eronen2025structural,
  title={Structural Sensitivity of N 1s Excitations in N-Methylacetamide Solutions},
  author={Eronen, E.A. and Vladyka, A. and Sahle, Ch. J. and Niskanen, J.},
  journal={The Journal of Physical Chemistry Letters},
  volume={16},
  number={7},
  pages={1666--1672},
  year={2025},
  publisher={ACS Publications}
}

@article{eronen2025statistics,
  title={Statistics within UV-Visible Absorption Spectrum of Ethanolic Azobenzene},
  author={Eronen, Eemeli A and Niskanen, Johannes},
  journal={arXiv preprint arXiv:2505.02610},
  year={2025}
}

@Article{Wikfeldt2009,
  author   = {Wikfeldt, Kjartan T. and Leetmaa, Mikael and Ljungberg, Mathias P. and Nilsson, Anders and Pettersson, Lars G. M.},
  journal  = {The Journal of Physical Chemistry B},
  title    = {On the Range of Water Structure Models Compatible with X-ray and Neutron Diffraction Data},
  year     = {2009},
  number   = {18},
  pages    = {6246-6255},
  volume   = {113}
}

@Article{Pettersson2022,
  author   = {Lars G.M. Pettersson and Osamu Takahashi},
  journal  = {Journal of Non-Crystalline Solids: X},
  title    = {The local structure of water from combining diffraction and X-ray spectroscopy},
  year     = {2022},
  issn     = {2590-1591},
  pages    = {100087},
  volume   = {14},
  doi      = {https://doi.org/10.1016/j.nocx.2022.100087},
}

@article{timoshenko2017supervised,
  title={Supervised machine-learning-based determination of three-dimensional structure of metallic nanoparticles},
  author={Timoshenko, Janis and Lu, Deyu and Lin, Yuewei and Frenkel, Anatoly I},
  journal={The journal of physical chemistry letters},
  volume={8},
  number={20},
  pages={5091--5098},
  year={2017},
  publisher={ACS Publications}
}

\onecolumngrid
\clearpage
\section*{Supplementary Information}

\subsection{Mean Integrated Absolute Deviation vs. Nominal Acceptance Rate in DRRMC}

\begin{figure}[htbp]
    \centering
    \includegraphics[width=.49\linewidth]{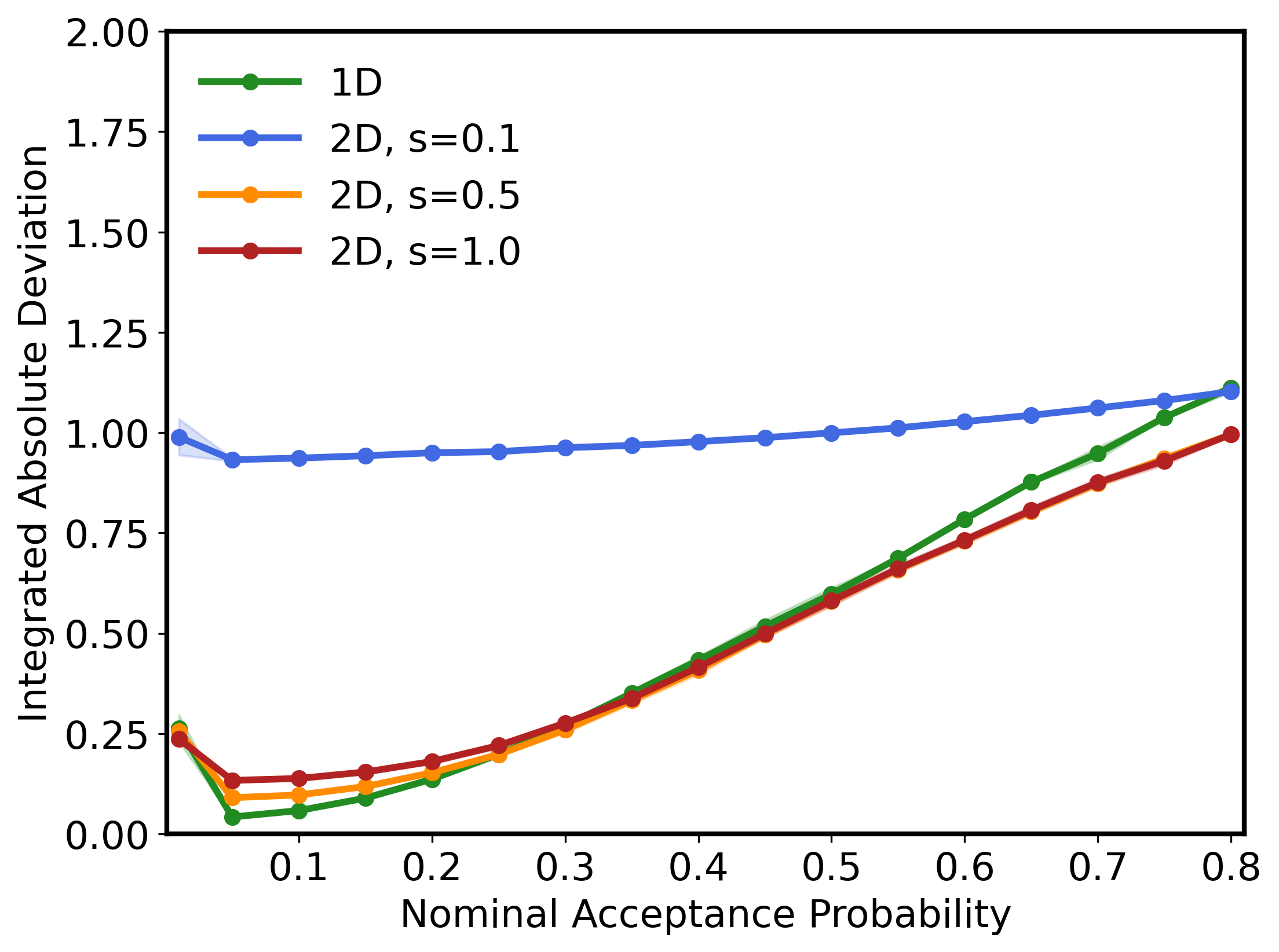}
    \hfill
    \includegraphics[width=.49\linewidth]{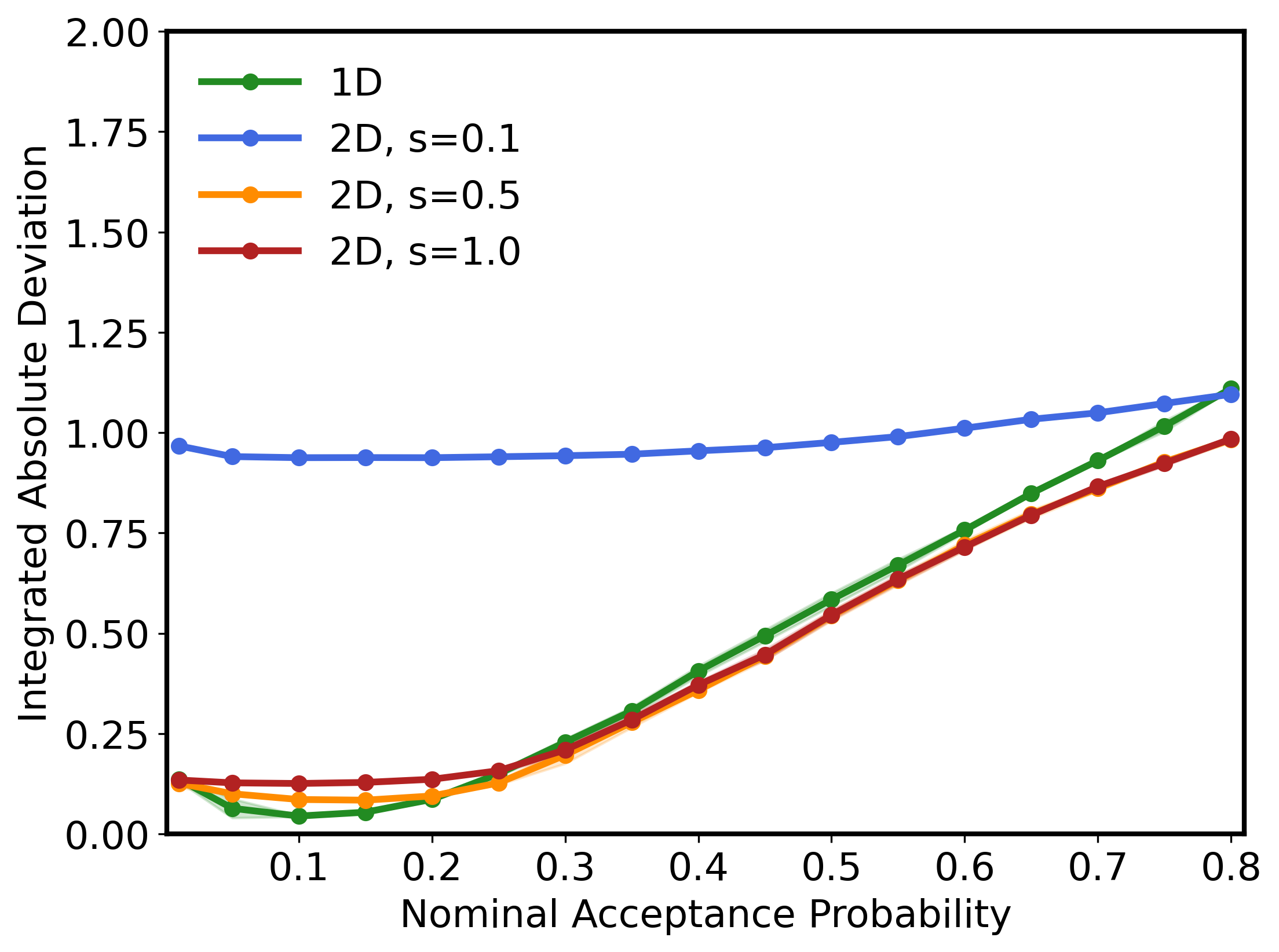}
    \caption{Mean integrated absolute deviation between the reconstructed and the reference
    distributions as a function of the nominal acceptance probability in the DRRMC
    for one- and two-dimensional cases. The left panel corresponds to sample size
    $N_s=64$, and the right panel to sample size $N_s=256$. Curves represent
    averages over 10 sampling runs with different random seeds, while the shaded
    regions denote $\pm1$ standard deviation.}
    \label{fig:SI_accept_rate}
\end{figure}

\clearpage

\subsection{Comparison of Reference and Reconstructed Two-Dimensional Distributions for XPS, XAS, and XES Spectra}

\begin{figure}[htbp]
    \centering
    \includegraphics[width=\linewidth]{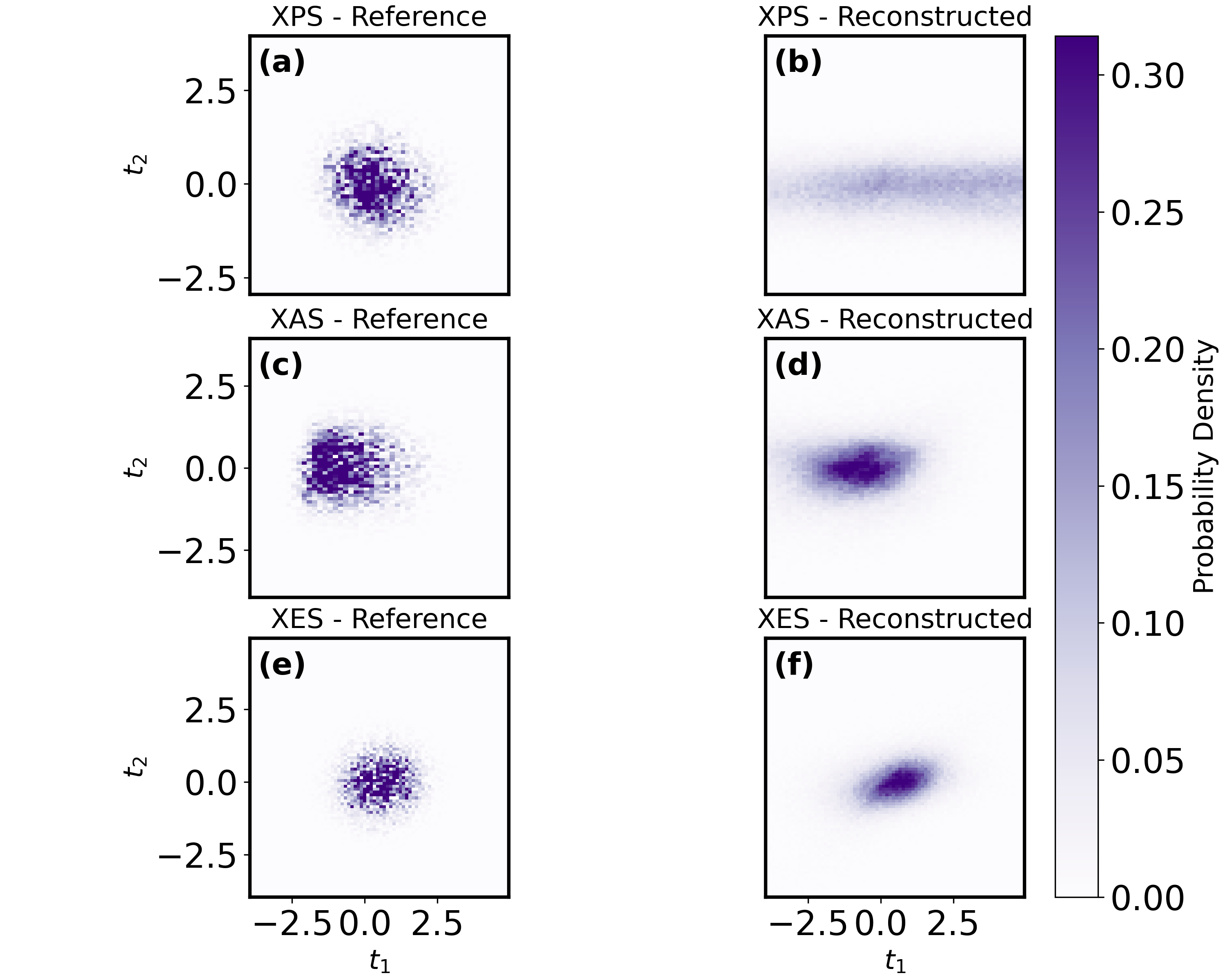}
    \caption{Two-dimensional distributions for XPS, XAS, and XES datasets. The columns
    show the reference distributions (left) and their corresponding reconstructions
    (right). All maps are normalized and displayed on a shared logarithmic color
    scale, enabling direct comparison of reconstructed structural distributions.
    Note that the basis vectors differ for different spectroscopies, which is
    manifested as shifts between the distributions.}
    \label{fig:SI_xps_xas_xes}
\end{figure}

\end{document}